\documentclass[12pt]{article}

\usepackage{amssymb}

\begin{document}

\begin{titlepage}

\begin{flushright}
IUHET-506\\
hep-ph/0703126
\end{flushright}
\vskip 2.5cm

\begin{center}
{\Large \bf Bound on the Photon Charge from the\\
Phase Coherence of Extragalactic Radiation}
\end{center}

\vspace{1ex}

\begin{center}
{\large Brett Altschul\footnote{{\tt baltschu@indiana.edu}}}

\vspace{5mm}
{\sl Department of Physics} \\
{\sl Indiana University} \\
{\sl Bloomington, IN 47405 USA} \\

\end{center}

\vspace{2.5ex}

\medskip

\centerline {\bf Abstract}

\bigskip

If the photon possessed a nonzero charge, then electromagnetic waves traveling
along different paths would acquire Aharonov-Bohm phase differences. The fact that
such an effect has not hindered interferometric astronomy places a bound on
the photon charge estimated to be at the $10^{-32}$ $e$ level if all photons have
the same charge and $10^{-46}$ $e$ if different photons can carry different charges.

\bigskip

\end{titlepage}

\newpage

In 2006, the Particle Data Group~\cite{ref-PDG} listed only four bounds on the photon
charge, compared with fifteen on the photon mass. The photon's electrical neutrality
and its masslessness are both quite important to our understanding of
electrodynamics.
We would therefore like to have the best bounds possible for both these quantities,
yet it seems that significantly more work has gone into constraining the
photon's mass. In some sense, this is unsurprising, since a nonzero photon mass is
much easier to accommodate theoretically. However, a significant improvement
over the currently quoted bounds on the photon charge is possible, if quantum
interference effects are considered

There are some crucial differences between a photon mass and a photon charge. A
photon mass parameter has meaning even at the classical level, where it determines
the gap in the dispersion relation. A photon charge, on the other hard, is
intrinsically quantum mechanical in nature. The charge of a propagating wave is
crucially tied to the decomposition of that wave as a collection of quantized
photons. Moreover, while there are at least three viable dynamical models for giving
mass to an Abelian gauge field (Proca, Higgs, and Stueckelberg), it is
much more difficult
to construct a consistent model of charged gauge bosons. The only straightforward
way to make gauge fields charged is to use a non-Abelian gauge group, but this
requires the existence of a nontrivial multiplet of vector bosons. Related to this
is the fact that electric charge is observed to be quantized in units of the proton
charge $e$ (in a way that mass is not); yet the present limits on the photon charge
are already many orders of magnitude smaller than $e$.

The absence of a complete theory describing charged photons can complicate the
task of placing bounds on any such putative charge. It is possible to place bounds
on the photon mass using dynamical stability conditions~\cite{ref-chibisov} and
observations of magnetohydrodynamic waves~\cite{ref-barnes}, as well as tests with
static fields~\cite{ref-davis}. These tests are possible because we can interpret
the results of our observations in the context of a well-defined theory---although
there may be unexpected model dependences in the conclusions~\cite{ref-adelberger}.
Without any equivalent theory describing the photon charge, we must rely on rather
different techniques. For example, while positing the existence of a photon charge
does tell us things about how photon propagation must be affected, it does not tell
us anything about how static electromagnetic fields will be modified.

Any bound placed on the photon charge is going to be subject to some level of
uncertainty, but some measurements are more robust that others. A very simple-minded
experiment involves observing the change in a photon's energy between a source and
detector placed at different voltages. Other measurements involve the deflection of
photons by magnetic fields~\cite{ref-cocconi1,ref-semertzidis,ref-kobychev} and the
associated time delays~\cite{ref-cocconi2,ref-raffelt}. The best trustworthy bound
derived by these methods is at the $4\times 10^{-31}$ $e$ level~\cite{ref-kobychev}.
(A previously quoted better bound in~\cite{ref-cocconi2} is considered to have been
in error.) Moreover, the bounds are actually two orders of magnitude better if both
positively and negatively charged photons are assumed to exist.

Indirect searches for the effects of a photon charge are also possible. Measurements
of the anisotropy of the cosmic microwave background (CMB) can be used to place
somewhat model-dependent bounds on the photon charge~\cite{ref-sivaram,ref-caprini}.
If there are
no cancellations between different species, and a nonzero photon charge disturbed the
overall charge neutrality of the early universe, there would be signatures in the
CMB. With these assumptions, a limit can be placed on the photon charge at the
$10^{-35}$ $e$ level.

We shall suggest a more subtle direct test, which is fundamentally quantum mechanical
and based on the Aharonov-Bohm effect. Charged particles moving along different paths
through a magnetic field
pick up different phases, and the observed coherence of photons from distant
astrophysical sources will allow us to place bounds on this effect and hence on the
photon charge.
%We shall not consider the possibility that different photons may
%carry different charges, because of the possibility of interference between them.
%In general, the existence of interference between differently charged particles would
%badly violate gauge invariance.
Other sensitive interferometric techniques, such as intensity interferometry, might
also be useful for setting bounds.

The bounds we shall place are based on observations of photons that have
traversed cosmological distances. These photons can be very precise probes of
novel phenomena in electrodynamics, the immense distances over which they
travel magnifying miniscule effects. A tiny change in how electromagnetic waves
propagate can, after millions of parsecs, give rise to a readily observable effect.
The strategy of observing radiation from very distant sources has already been
used to place extremely stringent bounds on photon birefringence~\cite{ref-carroll2,
ref-kost11,ref-kost21} and dispersion~\cite{ref-ellis}.

Our bounds should not
depend in any crucial way on the intricate details of the charged
photon dynamics. We shall assume only that there exists an effective Lagrangian
governing the propagation of a single photon, and that the coupling of the photon
to the external electromagnetic field takes the form $L_{I}=
-\frac{q}{c}v_{\mu}A^{\mu}_{{\rm ext}}$. The photon's charge is $q$, and
$v^{\mu}=(c,c\hat{v})$ is its
four-velocity. This Lagrangian is essentially unique, once we specify that there must
be a potential energy term $-qA^{0}$ and demand conventional Lorentz transformation
properties. The equation of motion derived from $L_{I}$ is the Lorentz force law.

From $L_{I}$, we can calculate the additional phase that a charged photon
picks up as it travels, relative to a conventional uncharged photon. In the eikonal
approximation, in which the photon's deflection from a straight-line path is
neglected, the phase is
\begin{equation}
\phi=\frac{1}{\hbar}\int_{0}^{t}d\tau\, {\cal L}_{I},
\end{equation}
where we have taken the time interval of the photon's flight to range from 0 to $t$.
Neglecting the contribution of the electrostatic potential and taking the total
distance traveled to be $L$, the phase is
\begin{equation}
\label{eq-phase}
\phi=\frac{q}{\hbar c}\int_{0}^{L}d\vec{\ell}\cdot\vec{A}_{{\rm ext}}.
\end{equation}

What we can observe is the phase difference between photons arriving at
different points. In practice, this is done all the time, and it is the basis of
astrophysical interferometry. We consider two telescopes, separated by a
baseline $d$. They collect data from a source lying approximately in the plane
perpendicular to the baseline. The observed phase difference due to a possible photon
charge is equal to the difference between two phases $\phi_{1}$ and $\phi_{2}$ of the
form (\ref{eq-phase}). Neglecting a miniscule contribution
proportional to the integral of $\vec{A}_{{\rm ext}}$ along the baseline, the
phase difference is $\Delta\phi=\Phi q/\hbar c$, where $\Phi$ is the magnetic flux
threading between the two lines of sight. This is the standard Aharonov-Bohm phase
difference, and it is independent of the photon energy.
%(Moreover, it is clear that
%if different photons carried different charges, their phase differences would depend
%explicitly on the vector potential $\vec{A}_{{\rm ext}}$, making it very difficult to
%develop a sensible interpretation of this phenomenon.)

To estimate the flux integral, we must know something about the relevant magnetic
fields. For randomly oriented fields, with typical magnitude $B$ and correlation
length $\lambda_{C}$, the flux $\Phi$ depends on $Bd\sqrt{\lambda_{C}L}$. To get
an idea of the accompanying numerical constant, we assume that the line of sight
passes though $L/\lambda_{C}$ magnetic field domains, each of equal size. In each
domain, the field is randomly oriented along one of six cardinal directions.
One third of the domains contribute to the total flux, behaving like a random walk.
The mean distance from the origin in this walk after $L/3\lambda_{C}$ steps
is $\sqrt{2L/3\pi\lambda_{C}}$, and with an extra factor of $\frac{1}{2}$
corresponding to the triangular geometry of the threaded region, the total flux
is
\begin{equation}
\Phi\approx\sqrt{\frac{L\lambda_{C}}{6\pi}}dB.
\end{equation}

An Aharonov-Bohm phase could upset interferometric measurements. In order for
interferometry to be possible, photons arriving at different telescopes must have
definite phase relations. A $\Delta\phi$ of order 1 would destroy this
necessary
relation. (While there are phase uncertainties in real measurements due to
uncertainties in telescope positions, these have very different characteristics and
can be distinguished from an Aharonov-Bohm phase.
Position uncertainties lead to phase differences that are proportional to the photon
frequency, and they have a predictable dependence on the direction of observation,
whereas the Aharonov-Bohm phase is frequency independent and varies randomly with
different pointing directions. Telescopes are calibrated by observing reference
sources, but as long as these are relatively nearby, the bounds will not be
significantly affected, since the Aharonov-Bohm phase is related to the distance;
there is no problem with this for the telescope arrangement discussed
below~\cite{ref-scott}.) So our ability to study objects
at a distance $L$ with interferometers of baseline $d$ limits the photon charge to
be smaller than
\begin{equation}
\label{eq-bound}
\frac{|q|}{e}<\sqrt{\frac{6\pi}{L\lambda_{C}}}\frac{\hbar c}{deB}.
\end{equation}

It is worthwhile to contrast this bound with the one that can be derived from
measurements of photon deflection by a magnetic field. That bound also depends on
the transverse magnetic field along the line of sight. For a constant field $B$
and line of sight $L$, with photons of energy $E$ and energy spread $\Delta E$,
the bound is formulated in terms of the angular deviation $\Delta\theta$ of the
different-energy photons as
\begin{equation}
\label{eq-deflect}
\frac{|q|}{e}<\frac{2E^{2}\Delta\theta}{ceBL\Delta E}.
\end{equation}
(If astrophysical field configurations are used, $BL$ would again be replaced by
something of order $B\sqrt{L\lambda_{C}}$.) It is also possible, when using photons
of single energy in the laboratory, to use a time-dependent magnetic field and then
look for a corresponding time-dependent angular deviation. In either case,
the deflection decreases linearly with the photon energy, since higher-energy
photons possess more momentum and are thus less deflected by the energy-independent
Lorentz force. The Aharonov-Bohm phase is independent of energy, although it is
still most advantageous to work with low-energy photons, because their phases can
be determined most accurately. The Aharonov-Bohm phase $\Delta\phi$ also depends
unavoidably on $\hbar$, while the bound (\ref{eq-deflect}) can evidently be
written in an $\hbar$-independent form. There is therefore no intrinsic relationship
between the bounds obtained by the two different methods, and it should be no
surprise if the experimental bounds available via the two methods differ by orders
of magnitude.

The greatest uncertainty in our photon charge
bounds will come simply from a limited understanding
of extragalactic magnetic fields. The best bounds on extragalactic fields come
from observations of the Faraday rotation of photons moving through putatively
magnetized plasmas~\cite{ref-vallee,ref-kronberg,ref-blase}. The precise bounds
one may derive from the Faraday observations depend on the assumptions one makes
about the large scale structure of the field and particularly on $\lambda_{C}$.
A bound of $B\lesssim10^{-8}$ G is reasonable, while cosmic ray and
high-energy photon data from the source Centaurus A suggest that $10^{-8}$ G may
also be a lower bound for the magnetic field strength in the relative vicinity of our
galaxy~\cite{ref-anchordoqui}.

To be conservative, we shall assume a rather lower value of the extragalactic
magnetic field. Cosmic ray data suggest that $B\sqrt{\lambda_{C}}$ may be at the
$10^{-10}$ G Mpc$^{1/2}$ level~\cite{ref-lemoine}, although
this still depends on assumptions about the distribution of ultra-high-energy cosmic
ray sources. A possible higher density of sources would yield a higher value of the
field.

$B\sqrt{\lambda_{C}}$ is an essentially universal (if somewhat uncertain) quantity,
but the parameters $d$ and $L$ that enter into (\ref{eq-bound}) are experimental
variables. The longest baseline $d$ available is that of the Very Long Baseline
Interferometry Space Observatory Program (VSOP) satellite experiment, for which
$d>3\times 10^9$ cm. Using radio telescopes on Earth in combination with one on
the Highly Advanced Laboratory for Communications and Astronomy satellite,
VSOP imaged active galactic nuclei out to redshifts $z>3$~\cite{ref-scott}. For
objects this
distant, the effects of cosmological expansion could not be ignored in any precise
treatment. However, for obtaining conservative order-of-magnitude bounds on $q$,
this level of precision is not overly important. Excellent bounds on the photon
charge can be derived merely from taking $L\sim 1$ Gpc, which is  roughly half an
order of magnitude smaller than the Hubble distance and corresponds to a redshift
less than 1.

Taking $B\sqrt{\lambda_{C}}=10^{-10}$ G Mpc$^{1/2}$, $d=3\times 10^9$ cm, and
$L=1$ Gpc, we find our bound on the photon charge to be
\begin{equation}
\label{eq-numeric}
\frac{|q|}{e}\lesssim 10^{-32},
\end{equation}
which is an improvement over all previous direct bounds that do not consider
photons with multiple opposing charges.
%The improvement over the trustworthy time delay bounds is more than two orders of
%magnitude.
There is even a small improvement relative to the erroneously stated bound
from~\cite{ref-cocconi2}.

As previously mentioned, it has also been suggested that there may exist photons with
positive and negative (or positive, negative, and zero) charges. If this is the case,
the CMB bounds would not apply, since the photon gas filling the early universe would
be charge neutral. The coherence of observed photons places bounds on this
possibility as well. However, there are additional uncertainties if particles with
different charges can interfere (and no pairs of
like-polarized photons have ever been observed not
to interfere). Although the phase difference for two particles of equal charge is
always gauge invariant, for particles with different charges it is not. Therefore,
the gauge in this scenario must be specified. It is no surprise that gauge invariance
is destroyed, since interference between dissimilarly charged particles violates
local charge conservation and charge superselection. The gauge fixing condition
should arise naturally in the full theory describing this phenomenon, just as the
Lorenz gauge condition $\partial^{\mu}A_{\mu}=0$ arises if we introduce a Proca mass
term. Yet without a complete theory, the precise form of the gauge condition is
unknown.

Nevertheless, it is possible to place an order of magnitude bound on the charge,
provided the large scale structure of $\vec{A}$ is not modified. In a magnetic field
domain, the typical vector potential is $B\lambda_{C}/2$. Conservatively assuming
that
the potential falls back to zero at the edge of the domain, the net contribution to
the phase for a photon of charge $q$ is $\phi=\sqrt{\frac{L}{6\pi}}\frac{\lambda_{C}
^{3/2}qB}{\hbar c}$, with the same factor of $\sqrt{2L/3\pi\lambda_{C}}$ as before.
The phase difference for photons of charges $q$ and $-q$ is twice this, even if the
photons traverse exactly the same path. Because of this, the bounds are improved by
a factor of ${\cal O}(d/\lambda_{C})$. $\lambda_{C}$ is more difficult to determine
than $B^{2}\lambda_{C}$, but choosing a relatively conservative value of 100 kpc
gives an improvement of ${\cal O}(10^{-14})$,
or a bound of
\begin{equation}
\label{eq-numstrong}
\frac{|q|}{e}\lesssim 10^{-46}
\end{equation}
if multiple charges are possible. This is a major improvement over previous bounds.

Moreover,
the bounds (\ref{eq-numeric}) and (\ref{eq-numstrong}) may still be relatively
conservative. They assumed
what might be too low a value for the extragalactic magnetic field and a length
scale $L$ corresponding to a relatively modest redshift. Perhaps most importantly,
the bound assumes that only a phase decoherence $\Delta\phi\sim 1$ is ruled out by
the availability of interferometric data. However, we feel that this level of
conservatism is warranted, given the uncertainties in quantities such as
$B\sqrt{\lambda_{C}}$.

Significant improvements in these kinds of bounds on the photon charge are possible,
but only up to a certain point.
A better understanding of magnetism on extragalactic scales will provide more
secure (but not necessarily numerically tighter) bounds on $q$. More careful
analyses, taking into account the expansion of the universe, could also extend the
reach in $L$, but since the dependence on $L$ is only as $L^{-1/2}$, the gain to
be had in this area is not great. Tighter limits on the experimentally observed phase
deviation $\Delta\phi$ would give proportionately tighter bounds on the charge.
The largest improvements in the single charge case might come from using longer
baselines. In
principle, a baseline of 2 AU could be possible for certain types of interferometric
measurements, and a baseline this long would improve the bound on $q$ by four
orders of magnitude.

The photons observed by the VSOP experiment had frequencies of 1.6, 5, and 22 GHz.
This places the energies of the photons from which our bound on $q$ was derived in
the 6--90 $\mu$eV range at the time of their absorption. We might expect, based on
Lorentz invariance and charge conservation, that the photon charge should be
independent of energy, as is the charge of other particles; however, more exotic
possibilities cannot be ruled out.

A critic
might object that the Aharonov-Bohm-type phase should not contribute to the
ordinary, essentially classical, phase that we observe in radio waves. In this view,
the novel phase would represent some kind of intrinsically quantum effect, one which
could be observed only if a single photon whose wave packet had been split were
recombined and then observed (as opposed to observing distinct but coherent photons
at different locations). The corresponding interference effects would have to
operate entirely separately from the usual interference of electromagnetic waves.
Without a viable theory of charged photons, we cannot definitively reject such a
hypothesis; we do, however, note that it violates the usual correspondence principle
relationship that connects photons with classical waves, in which the classical and
photon phases are one and the same.

We have presented improved bounds on the magnitude of the photon charge, derived from
quantum interference considerations. Charged photons, like other charged particles,
could acquire Aharonov-Bohm phases, yet no evidence for these phases has been seen
in interferometry experiments. Using emissions from distant astrophysical
sources---which are particularly useful for constraining small deviations from
conventional electrodynamics---the fraction of the fundamental charge present on a
radio-frequency photon has been bounded at the $10^{-32}$ or $10^{-46}$
levels, depending on whether all photons carry the same charges and provided the
large scale structure of the vector potential is unmodified.

\section*{Acknowledgments}
The author is grateful to V. V. Kobychev for his helpful comments.
This work is supported in part by funds provided by the U. S.
Department of Energy (D.O.E.) under cooperative research agreement
DE-FG02-91ER40661.


\begin{thebibliography}{99}

\bibitem{ref-PDG}W.-M. Yao, et al., J. Phys. G {\bf 33}, 1 (2006).
\bibitem{ref-chibisov}G. Chibisov, Usp. Fiz. Nauk {\bf 19}, 551 (1976); Sov. Phys.
Usp. {\bf 19}, 624 (1976).
\bibitem{ref-barnes}A. Barnes, J. D. Scargle, Phys. Rev. Lett. {\bf 35}, 1117
(1975).
\bibitem{ref-davis}L. Davis, A. S. Goldhaber, M. M. Nieto, Phys. Rev. Lett.
{\bf 35}, 1402 (1975).
\bibitem{ref-adelberger}E. Adelberger, G. Dvali, A. Gruzinov, Phys. Rev. Lett. 98,
010402 (2007).
\bibitem{ref-cocconi1}G. Cocconi, Amer. J. Phys. {\bf 60}, 750 (1992).
\bibitem{ref-semertzidis}Y. K. Semertzidis, G. T. Danby, D. M. Lazarus, Phys. Rev. D
{\bf 67}, 017701 (2003).
\bibitem{ref-kobychev}V. V. Kobychev, S. B. Popov, Astron. Lett. {\bf 31}, 147
(2005).
\bibitem{ref-cocconi2}G. Cocconi, Phys. Lett. B {\bf 206}, 705 (1988).
\bibitem{ref-raffelt}G. Raffelt, Phys. Rev. D {\bf 50}, 7729 (1994).
\bibitem{ref-sivaram}C. Sivaram, Amer. J. Phys. {\bf 63}, 473 (1995).
\bibitem{ref-caprini}C. Caprini, P. G. Ferreira, JCAP {\bf 02}, 006 (2005).
\bibitem{ref-carroll2}S. M. Carroll, G. B. Field, Phys. Rev. Lett. {\bf 79}, 2394
(1997).
\bibitem{ref-kost11}V. A. Kosteleck\'{y}, M. Mewes, Phys. Rev. Lett. {\bf 87}, 251304
(2001).
\bibitem{ref-kost21}V. A. Kosteleck\'{y}, M. Mewes, Phys. Rev. Lett. {\bf 97}, 140401
(2006).
\bibitem{ref-ellis}J. Ellis, N. E. Mavromatos, D. V. Nanopoulos, A. S. Sakharov, E.
K. G. Sarkisyan, Astropart. Phys. {\bf 25}, 402 (2006).
\bibitem{ref-scott}W. K. Scott, {\em et al.}, Ap. J. Supp. {\bf 155}, 33 (2004).
\bibitem{ref-vallee}J. P. Vall\'{e}e, Astrophys. J. {\bf 360}, 1 (1990).
\bibitem{ref-kronberg}P. P. Kronberg, Rep. Prog. Phys. {\bf 57}, 325 (1994).
\bibitem{ref-blase}P. Blasi, S. Burles, A. V. Olinto, Astrophys. J. {\bf 514}, L79
(1999).
\bibitem{ref-anchordoqui}L. A. Anchordoqui, H. Goldberg, Phys. Rev. D {\bf 65},
021302 (R) (2001).
\bibitem{ref-lemoine}M. Lemoine, Phys. Rev. D {\bf 71}, 083007 (2005).

\end{thebibliography}
\end{document}